\documentclass[a4paper,fleqn]{cas-dc}

\usepackage[numbers]{natbib}

\def\tsc#1{\csdef{#1}{\textsc{\lowercase{#1}}\xspace}}
\tsc{WGM}
\tsc{QE}

\begin{document}
\let\WriteBookmarks\relax
\def\floatpagepagefraction{1}
\def\textpagefraction{.001}
\let\printorcid\relax 

\shorttitle{Nd3+ Doping-induced Leakage Currents Suppression in High-temperature 0.7BiFeO3-0.3BaTiO3 Lead-free Piezoceramics}    

\shortauthors{Zhengbao Yang et al.}

\title[mode = title]{Nd3+ Doping-induced Leakage Currents Suppression in High-temperature 0.7BiFeO3-0.3BaTiO3 Lead-free Piezoceramics}

\author[1]{Jinming Liu}

\author[1]{Mingtong Chen}

\author[1]{Zhengbao Yang}
\cormark[1] 
\ead{zbyang@hk.ust} 
\ead[URL]{https://yanglab.hkust.edu.hk/}

\address[1]{The Hong Kong University of Science and Technology
Hong Kong, SAR 999077, China}

\cortext[1]{Corresponding author} 

\begin{abstract}
BiFeO\textsubscript{3} has attracted much attention as a potential candidate for replacing conventional, lead based piezoelectric materials due to its remarkable spontaneous polarization and high Curie temperature. However, its inherent high leakage currents, which lead to low piezoelectric response and poor temperature stability, have severely limited its practical applications. In this study, lead free piezoelectric ceramics of the 0.7BiFeO\textsubscript{3}-0.3BaTiO\textsubscript{3} (BF-BT) system were prepared, and their microstructures along with high-temperature electrical performance were modulated by introducing Nd\textsuperscript{3+}. The results indicate that moderate Nd doping improves lattice symmetry and reduces oxygen vacancy-related defect dipoles, thereby effectively suppressing leakage currents at temperatures above 75°C. The Nd-doped samples exhibit significantly lower leakage current densities, reduced by over 99\% compared to the undoped ceramics, reaching values as low as 10\textsuperscript{-5}A·cm\textsuperscript{-2}. They also show higher resistivity under elevated temperatures and electric fields, offering notable improvements in thermal stability over the undoped counterparts. In addition, the Nd-doped samples achieved piezoelectric coefficients as high as 172 pC N \textsuperscript{-1} at room temperature while still maintaining high dielectric and piezoelectric responses at elevated temperatures. This work not only provides an effective way to solve the leakage current problem of BF-BT ceramics in high temperature applications but also indicates a new design strategy for optimizing the high temperature stability of lead free piezoelectric materials, which shows a broad application prospect in the field of high-temperature sensors and actuators.

\end{abstract}




\maketitle

\section{Introduction}

Piezoelectric ceramics are widely used in actuators as miniature energy conversion elements to control displacements. Due to their fast, precise positioning control over nanometer micron scales, traditional Pb (Zr, Ti) O\textsubscript{3} (PZT) based ferroelectric materials have dominated the piezoelectric market for the last 60 years\cite{1,2}. However, because they contain lead, which is toxic and environmentally unfriendly, alternative lead free materials are urgently needed. Lead free piezoelectric ceramics, such as BaTiO\textsubscript{3} (BT)-based, BiFeO\textsubscript{3} (BF)-based, (K, Na) NbO\textsubscript{3} (KNN)-based, and (Bi, Na) TiO\textsubscript{3} (BNT)-based have therefore been extensively studied in recent years\cite{3,4,5,6}.  The comparison table is given in the Table 1, which shows the piezoelectric performance and Curie Temperature of the lead-free or lead-based piezoceramics.

\begin{table}[h]
\caption{d33  and Curie Temperature (Tc) of typical piezoceramics}\label{tbl1}
\begin{tabular*}{\tblwidth}{cccc}
\toprule
 Material & d33(pC N \textsuperscript{-1}) & Tc (°C) & Reference \\ 
\midrule
PZT-5H & 600-700 & ~200 & \cite{7} \\
BTO & 190 & 115 & \cite{8} \\
BFO & 27 & 830 & \cite{9} \\
BF-BT & 225 & 503 & \cite{10} \\
KNN & 500 & ~200 & \cite{11} \\
BNT & 98 & 315 & \cite{12} \\
\bottomrule
\end{tabular*}
\end{table}

BF-based piezoceramics exhibit significant ferroelectric and piezoelectric responses in a pseudocubic structure without extrinsic contributions from ferroelectric-domain dynamics. A giant electric-field induced strain of 5\% has been reported in BF thin films, attributed to the coexistence of tetragonal-like and rhombohedral-like phases and a field induced phase transition.\cite{13} In particular, BF BT ceramics have attracted interest as lead free alternatives due to their high piezoelectric coefficients and Curie temperature. They exhibit strong ferroelectric and piezoelectric responses in a pseudocubic matrix, challenging conventional understanding of piezoelectricity in cubic symmetry.\cite{14} Moreover, introducing BaTiO\textsubscript{3} to form BF-BT solid solutions greatly enhances the piezoelectric response near the morphotropic phase boundary (MPB) at x=0.7, where pseudocubic and rhombohedral phases coexist.\cite{15}

Although BF-BT normally has a high Curie temperature, at elevated temperatures and voltages, leakage currents are observed, which can cause ceramic breakdown. \cite{16} In an ideal capacitor there should be no DC leakage, but in practice leakage currents during DC operation heat the ceramic and shorten its lifetime. For ultra-low frequency drives, leakage currents also degrade positioning accuracy, so it is necessary to limit their magnitude. \cite{17,18}

Methods to limit leakage in BF-BT based ceramics include composition optimization and doping with Ga\textsuperscript{3+}, Co\textsuperscript{3+} or Sc\textsuperscript{3+} to improve resistivity. Table 2 summarizes recent results on piezoelectric performance and leakage suppression in doped BF-BT ceramics.

\begin{table*}[h]
\caption{Piezoelectric performance and leakage suppression of various BF-BT ceramics}\label{tbl2}
\begin{tabular*}{\tblwidth}{cccccc}
\toprule
 Material & d33(pC N \textsuperscript{-1}) & Temperature (°C) & Drive Field (kV·cm-1) & Current Density (A·cm-2) & Reference \\ 
\midrule
BF-BT,Co\textsuperscript{3+} & 151 & RT & 20 & 10\textsuperscript{-5} & \cite{19} \\
BF-BT,Ga\textsuperscript{3+} & 174 & RT & 10 & 10\textsuperscript{-8} & \cite{20} \\
BF-BT,Sc\textsuperscript{3+} & 165 & RT & 10 & 10\textsuperscript{-7} & \cite{21} \\
BF-BT & 178 & RT & 10 & 10\textsuperscript{-5} & This work \\
BF-BT & 238 & 125 & 10 & 10\textsuperscript{-3} & This work \\
BF-BT,Nd\textsuperscript{3+} & 227 & 125 & 10 & 10\textsuperscript{-5} & This work \\
\bottomrule
\end{tabular*}
\end{table*}

However, the effectiveness of these leakage suppression strategies is often reported only at room temperature and depends strongly on the applied field (e.g. Ga\textsuperscript{3+} doped ceramic measured at 2 kV·cm\textsubscript{-1} show J=10\textsuperscript{-9} A·cm\textsuperscript{-2} and rarely include temperature data). Considering many piezoelectric devices must operate in high temperature environments, studies of leakage currents in doped BF BT at elevated temperatures are scarce. The present work therefore focuses on suppressing leakage currents at elevated temperatures via Nd\textsuperscript{3+} doping in 0.7BF-0.3BT ceramics.

\section{Experimental Procedure}

A series of 0.7BiFeO\textsubscript{3}-0.3Ba\textsubscript{1-x}Nd\textsubscript{x}TiO\textsubscript{3}(x=0,0.005,\\
0.010,0.015) ceramics were fabricated via a conventional solid-state reaction method. Raw powders Bi2O3, Nd2O3, BaCO3 (>99\%, Sinopharm Chemical Reagent Co., Ltd), Fe2O3 (>99\%, Tianjin Fuchen), and TiO2 (>99\%, Aladdin) were purchased. The powders were dried at 110 °C for 24 h to remove moisture, then loaded into a rotating polyethylene jar with milling balls and dispersed in ethanol. The slurry was milled at 240 rpm for 24 h and then dried at 40 °C.Calcination was performed at 780°C and 800°C (4h each) to complete the solid‐state reaction. After recalcination, the powders were mixed with a polyvinyl alcohol (PVA) binder and sieved to ~100µm. Approximately 0.7g of powder was pressed into 11mm diameter disks, which were then heated at 600°C to burn out the binder. Sintering was carried out at 1020°C and 1040°C for 3h, followed by air quenching from 800°C. The disks were polished on both sides, coated with sputtered silver electrodes (less than1mm thickness for each sample), and annealed. Final poling was performed at 5 kV mm\textsuperscript{-1} in an oil bath at 70°C for 15min.

Ferroelectric polarization and electric field induced strain were measured using a CPE1801 ferroelectric tester (PolyK Technologies) equipped with a photonic sensor (MTI2100, MTI Instruments). Piezoelectric charge coefficients at various temperatures were tested with a quasi static d33 meter (TZFD 600, Harbin Julang Technology) at a heating rate of 3°C·min\textsuperscript{-1}.

Crystal structures were examined by X ray diffraction (D8 Advance, Bruker). In situ high temperature X ray scans were performed by heating at 20°C·min\textsuperscript{-1}, holding for 10min, then scanning; lattice parameters and phase content were refined using TOPAS3.0. Ferroelectric domains were observed by atomic force microscopy (Dimension Icon with ScanAsyst, Bruker) in optimized vertical deflection mode, using an Asyelec.01-R2 conductive probe (Asylum Research).

\section{Results and Discussion}

Figure 1 a, b show the SEM images of the cross section for BF-0.3BT ceramics at different Tsin. All the ceramics between 1020 °C and 1040 °C exhibit a clear grain boundaries. As x increases from 0.00 to 0.15, Nd3+ incorporation effectively refines the microstructure. By SEM, it is noted that Tsin has a varied influence on grain size.

As shown in Figure 2, at room temperature the undoped sample exhibits a d33 of 178 pC N \textsuperscript{-1}. From room temperature to 100 °C, d33 gradually increases as domain‐wall mobility is activated and the material transitions from a relaxation to a ferroelectric state\cite{22}. Between 100 °C and 200 °C, the d33 of the undoped sample remains nearly constant, forming a broad plateau with a peak value slightly below 250 pC N\textsuperscript{-1}. In contrast, above 200 °C, three doped samples rise again, peaking at 280 pC N\textsuperscript{-1} at 310 °C. Notably, the x = 0.005 sample attains 344 pC N\textsuperscript{-1} at 400 °C before sharply falling to zero above 400 °C, demonstrating markedly superior thermal stability compared to the undoped ceramic. Similar behavior has been reported in Sm\textsuperscript{3+} doped BF-BT ceramics\cite{23}.

\begin{figure*}[h]
	\centering
		\includegraphics[scale=0.85]{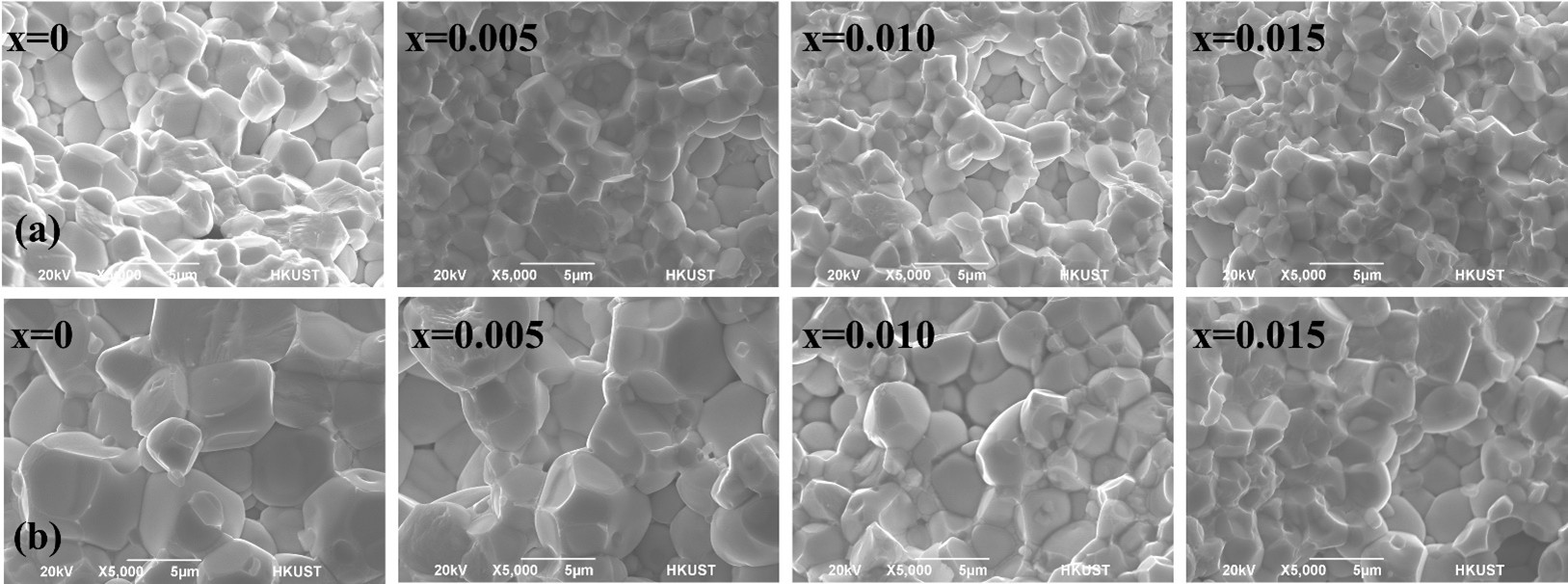}
	  \caption{SEM images of BF–0.3B1-xNdxT ceramics. a sintering temperature Tsin = 1020°C. b sintering temperature Tsin = 1040°C.}\label{FIG:1}
\end{figure*}

\begin{figure}[h]
	\centering
		\includegraphics[scale=0.85]{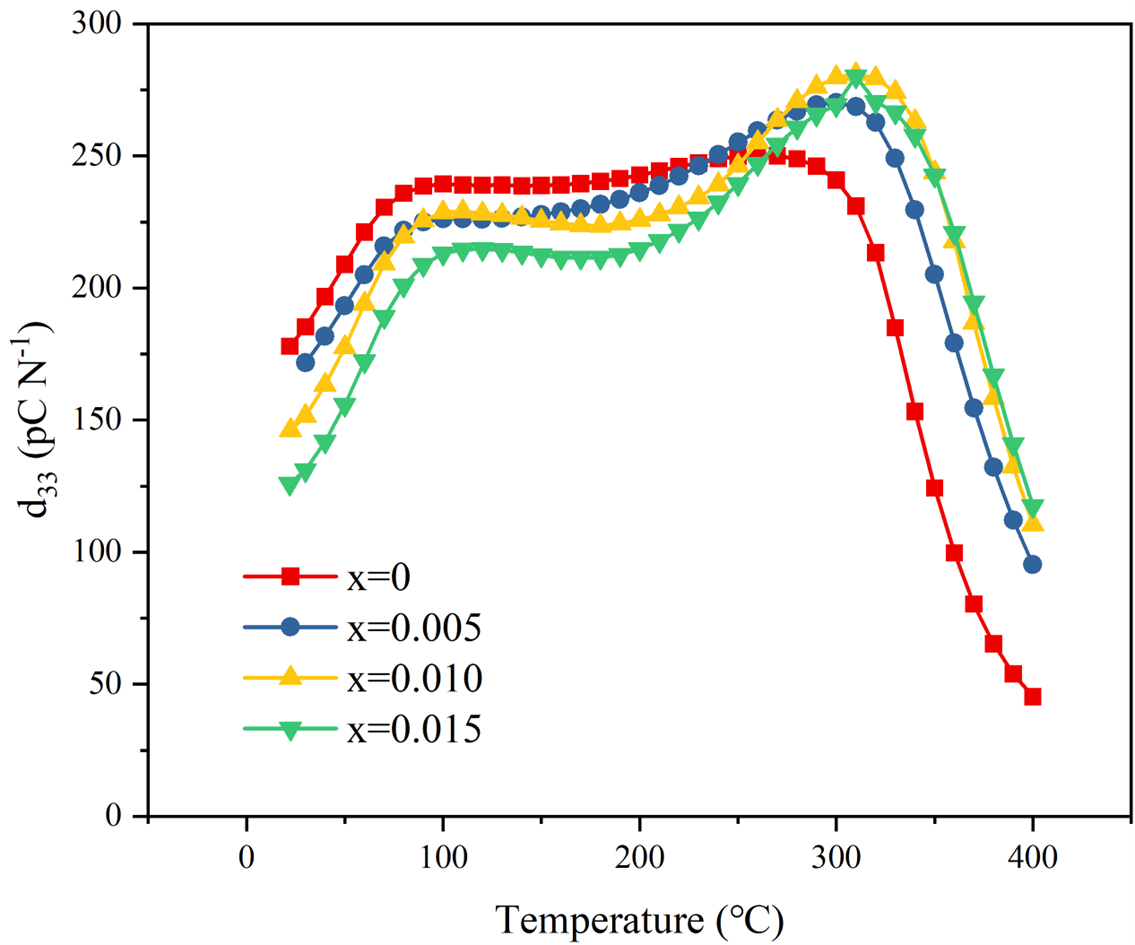}
	  \caption{In-situ temperature dependence of d33 for BF–0.3B\textsubscript{1-x}Nd\textsubscript{x}T ceramics with four Nd\textsuperscript{3+} contents (x = 0., 0.005, 0.010, 0.015).}\label{FIG:2}
\end{figure}

\begin{figure}[h]
	\centering
		\includegraphics[scale=0.85]{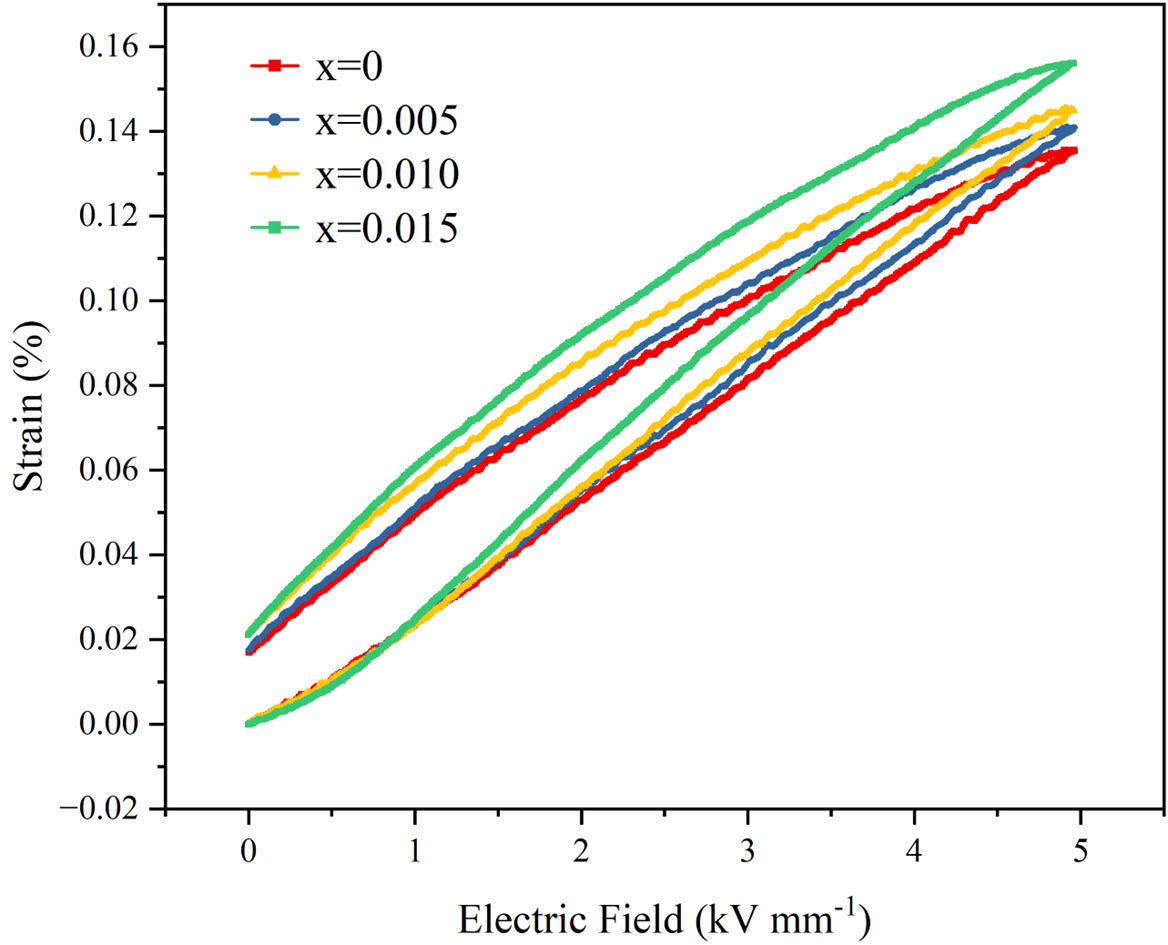}
	  \caption{Unipolar strain electric field (S E) curves at room temperature of BF–0.3B\textsubscript{1-x}Nd\textsubscript{x}T ceramics with four Nd\textsuperscript{3+}contents (x = 0.00, 0.005, 0.010, 0.015)}\label{FIG:3}
\end{figure}

The unipolar strain of BF-0.3B\textsubscript(1-x)Nd\textsubscript(x)T ceramics under different electric ﬁelds at room temperature is given in Figure 3. The maximum unipolar strain of 0.145 \% is observed in the x = 0.010 sample at 5 kV·mm\textsuperscript{-1}. Strain increases with increasing Nd content, suggesting lattice destabilization upon doping.

\begin{figure*}[h]
	\centering
		\includegraphics[scale=0.85]{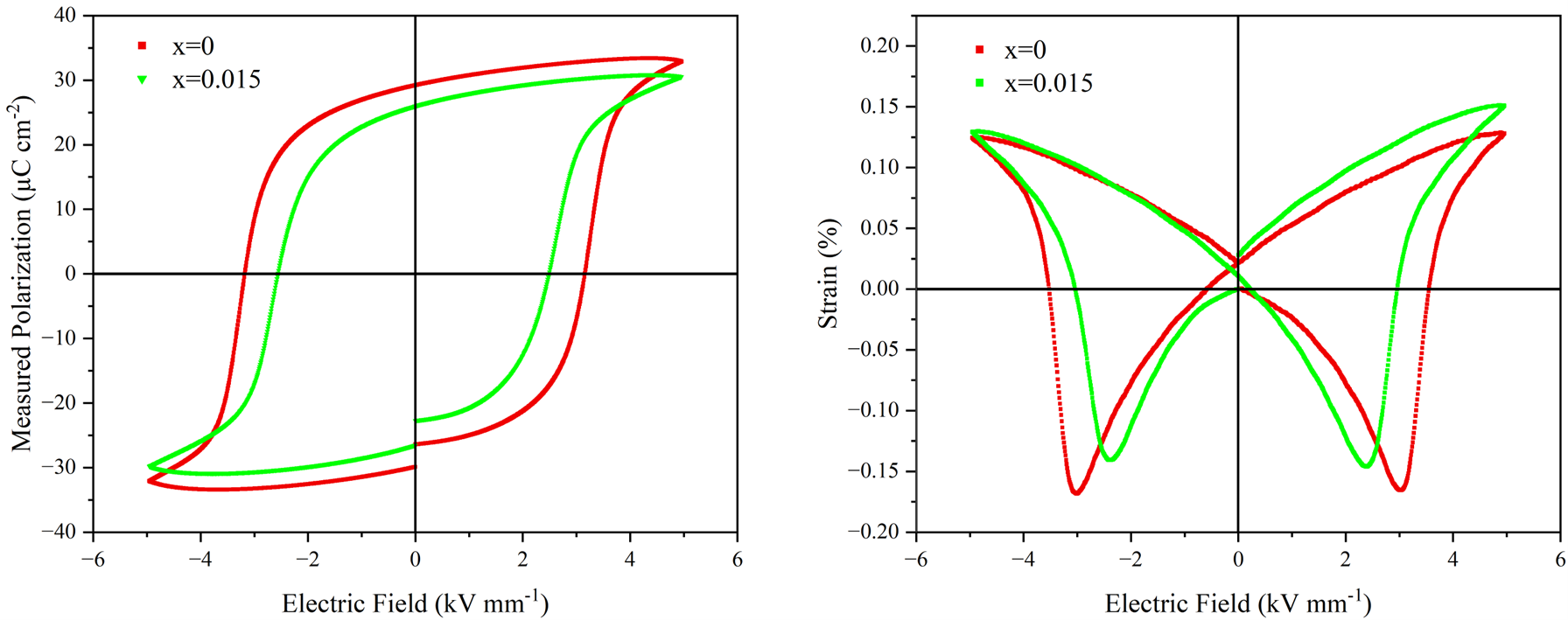}
	  \caption{The polarization (a) and strain (b) of undoped and Nd-doped samples under the bipolar electric field at 1 Hz}\label{FIG:4}
\end{figure*}

Figure 4a shows the polarization-electric field (P E) hysteresis loops of the poled samples. After Nd3+ doping, the P-E loop constricts, with reduced coercive field (E\textsubscript{0}) and remnant polarization (P\textsubscript{r}), yet still retains a high P of 26 µC cm-2. Figure 4b shows the strain–electric field (S-E) loops, the asymmetric shape is because that the sample was pre-poled with a DC field before the measurement, causing the alignment of defect dipoles. \cite{24,25}.The larger strain on the positive electric field is due to the initially applied electric field being same to the polarization direction of the sample. A decrease in coercive field and a significant increase in strain for the doped sample can also be observed.

\begin{figure*}[h]
	\centering
		\includegraphics[scale=0.85]{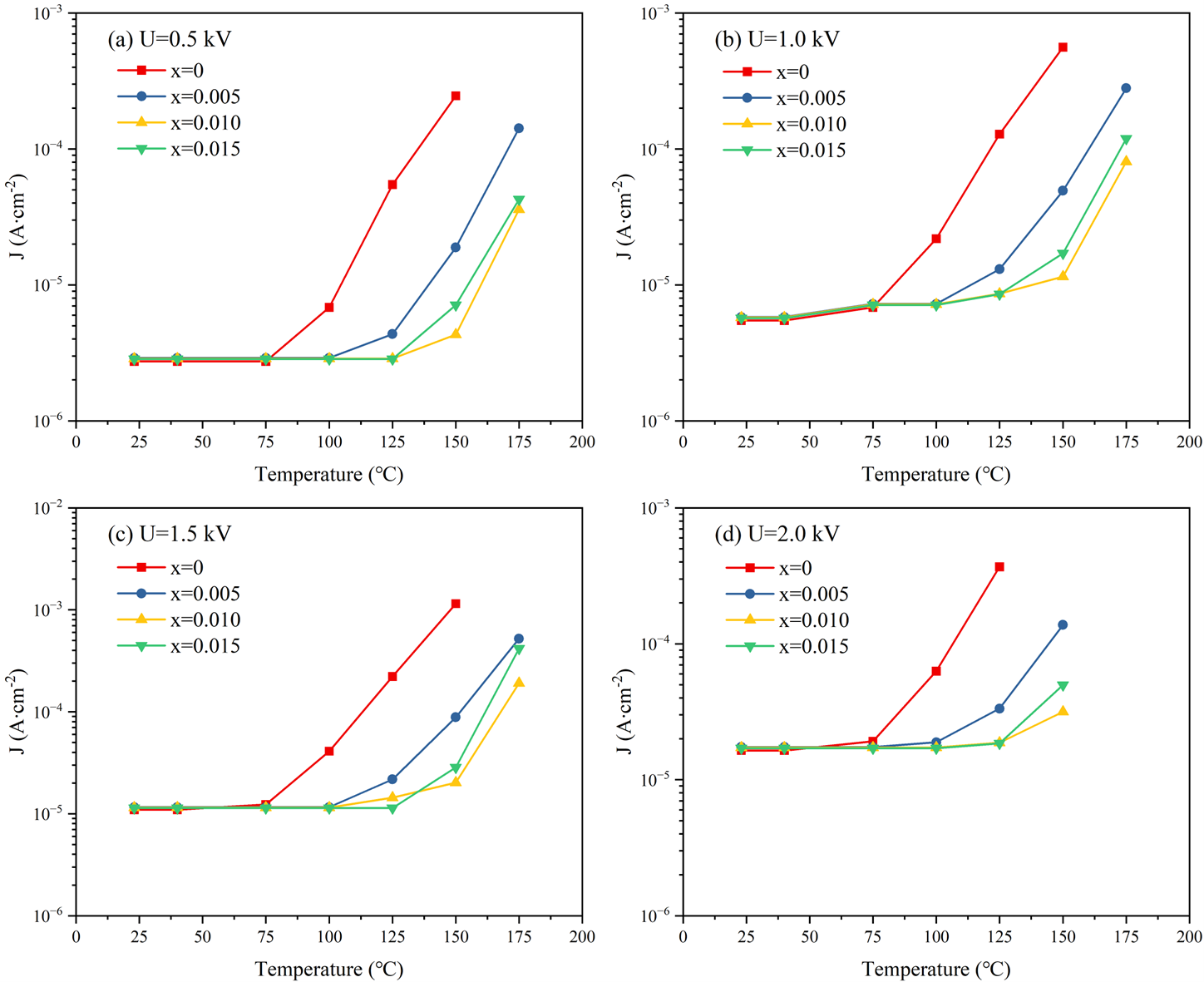}
	  \caption{The J -T curve for BF–0.3B\textsubscript{1-x}Nd\textsubscript{x}T ceramics: x = 0; 0.005; 0.010; 0.015 under 4 kinds of DC voltage. a U = 0.5 kV. b U = 1.0 kV. c U = 1.5 kV. d U = 2.0 kV.}\label{FIG:5}
\end{figure*}

Figures 5 a-d show the Current Density J - Temperature curve for BF–0.3B\textsubscript{1-x}Nd\textsubscript{x}T ceramics: x = 0.00; 0.005; 0.010; 0.015 under 4 kinds of DC voltage U = 0.5 kV (a), 1.0 kV (b), 1.5 kV (c), 2.0 kV (d) and different temperature ranges (RT, 40 ℃, 75 ℃, 100 ℃, 125 ℃, 150 ℃, 175 ℃). As the voltage increases from a-d, the value of leakage current increases. Under same voltage below 75 °C, all samples show comparable J at each bias. 
Above 75 °C, J of the undoped ceramics rises rapidly undergoing breakdown above 150 °C, whereas Nd\textsuperscript{3+} doped samples maintain lower J through higher temperatures. Thermal stability peaks at x = 0.010, with further Nd\textsuperscript{3+} addition causing a slight stability decline.
Supplement 1 a-d show the resistivity and temperature curve for BF–0.3B\textsubscript{1-x}Nd\textsubscript{x}T ceramics: x = 0; 0.005; 0.010; 0.015 under 4 kinds $\omega$ of DC voltage U = 0.5 kV (a), 1.0 kV (b), 1.5 kV (c), 2.0 kV (d). As temperature increases, $\rho$ decreases for all samples. However, Nd\textsuperscript{3+} doped ceramics consistently exhibit higher  than the undoped counterpart at each bias, confirming improved insulation and thermal stability.

The curve of resistivity versus temperature change is shown in Figure 6, but this time the temperature range is enlarged up to 400 ℃ with stepped voltage (100 V, 10 V, 1 V). As temperature rises to 275 ℃, the voltage is 10 V, and under 400 ℃ the voltage is 1 V. At room temperature, the undoped sample has $\rho$ = 9.6×10\textsuperscript{11} $\omega$·cm, whereas the x = 0.010 doped sample shows $\rho$ = 1.7×10$\rho$\textsuperscript{12} $\omega$·cm. On a logarithmic scale, this resistivity gap persists up to 400 °C, demonstrating the superior thermal resilience of Nd\textsuperscript{3+} doped BF BT ceramics.

\begin{figure}[h]
	\centering
		\includegraphics[scale=0.85]{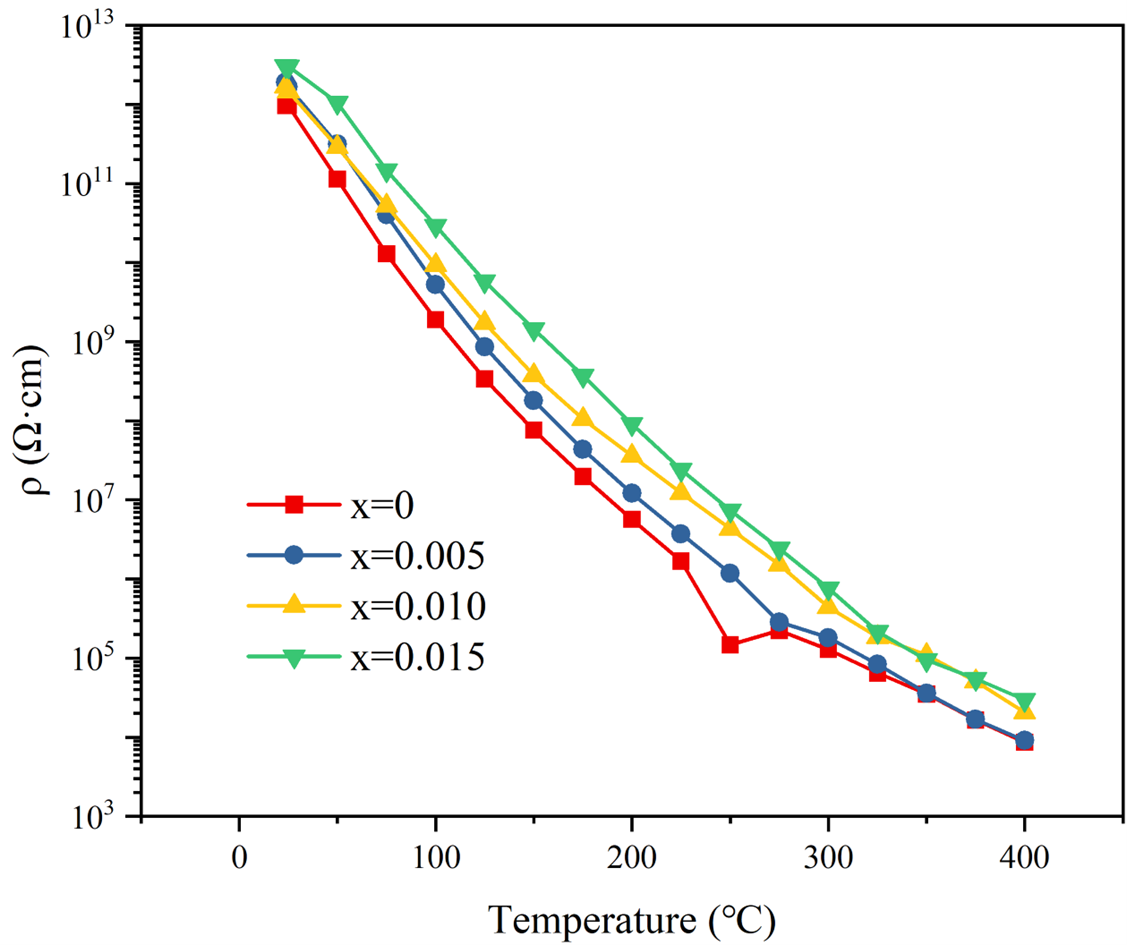}
	  \caption{The $\rho$-T curve for BF–0.3B\textsubscript{1-x}Nd\textsubscript{x}T ceramics: x = 0; 0.005; 0.010; 0.015 under stepped voltage (100V, 10V, 1V).}\label{FIG:6}
\end{figure}

\begin{figure}[h]
	\centering
		\includegraphics[scale=0.85]{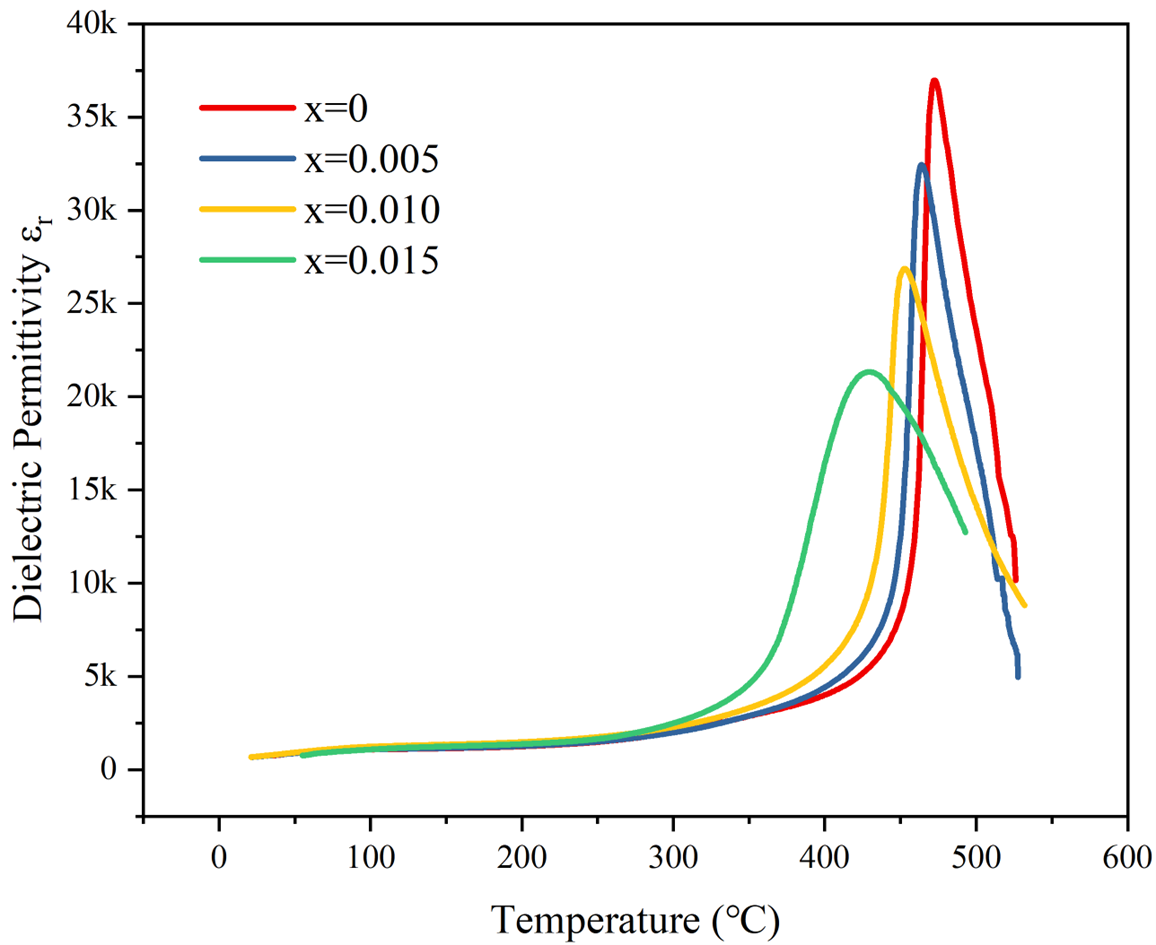}
	  \caption{Temperature dependence of $\epsilon$r at 10 kHz for BF–0.3B\textsubscript{1-x}Nd\textsubscript{x}T ceramics: x = 0; 0.005; 0.010; 0.015.}\label{FIG:7}
\end{figure}

As shown in the figure 7, the dielectric permittivity at various temperature exhibit a clear dependence on Nd doping content. For the undoped sample, a sharp and symmetric peak is observed at a relatively high temperature, demonstrating strong polarization coupling and clear Curie transition\cite{26}. With increasing Nd content, both the peak temperature and the maximum dielectric permittivity gradually decrease. It is noteworthy that when x = 0.010, the curve near the peak becomes wider, implying the occurrence of relaxation-like behavior. These changes can be explained by the substitution of Ba\textsuperscript{2+} by less polarizable Nd\textsuperscript{3+} ions, which weakens the polar interaction.\cite{27} Additionally, the chemical heterogeneity and local electric field fluctuations lead to the destruction of long-range ferroelectric order and the development of local polar nanodomains.\cite{28,29}

\begin{figure*}[h]
	\centering
		\includegraphics[scale=0.85]{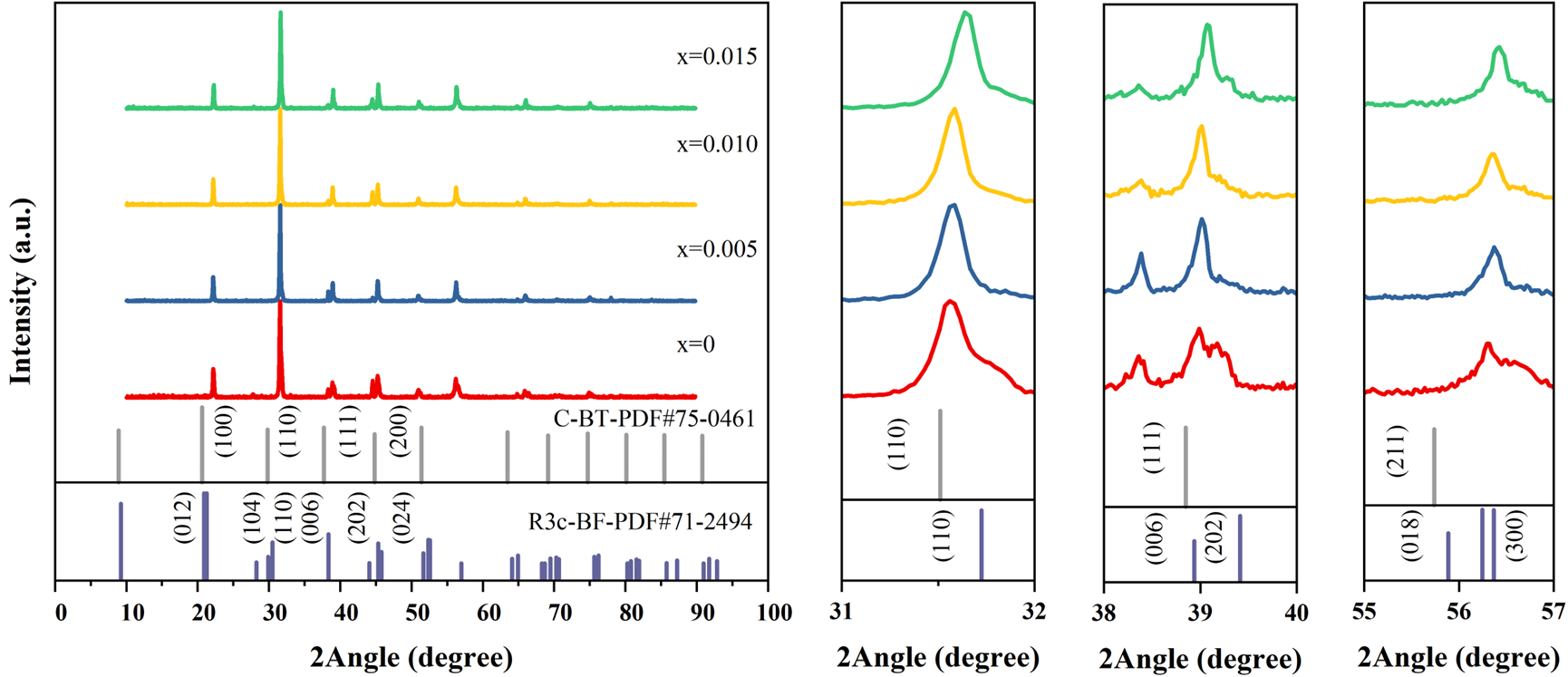}
	  \caption{The crystal structure. a X-ray diffraction patterns of BF–0.3B\textsubscript{1-x}Nd\textsubscript{x}T ceramic powder: x = 0; 0.005; 0.010; 0.015 at room temperature. b The local patterns around the (111)pc and (211)pc Bragg peaks.}\label{FIG:8}
\end{figure*}

Figure 8 presents the XRD patterns of BF–0.3B\textsubscript{1-x}Nd\textsubscript{x}T ceramics. For comparison, standard diffraction peaks of BiFeO\textsubscript{3} with R3c symmetry (PDF\#71-2494) and cubic BaTiO\textsubscript{3} (PDF\#75-0461) are indicated below. All samples exhibit a pure perovskite phase, and no diffraction peaks corresponding to Nd\textsubscript{2}O\textsubscript{3} are detected, suggesting that Nd\textsuperscript{3+} ions have been fully incorporated into the perovskite host lattice. Compared to the undoped sample, the diffraction peaks of the doped one shift toward higher angles, indicating a reduction in lattice volume, which can be attributed to the smaller ionic radius of Nd\textsuperscript{3+} (1.109 Å) compared to Ba\textsuperscript{2+} (1.61 Å).\cite{30}
In addition, the reduced splitting of certain diffraction peaks after doping implies an increase in crystal symmetry. Previous studies have shown that differences in ionic valence and radius can lead to structural and polarization discontinuities, raising the interfacial energy. \cite{31} The competition between bulk and interfacial energies may flatten the free-energy landscape, promoting a more symmetric structure with smaller domain size. Moreover, the higher electronegativity of Nd\textsuperscript{3+} compared to Ba\textsuperscript{2+} enhances local covalent bonding, and the hybridization of ionic and covalent bonds is considered beneficial for improving the thermal stability.
In the BF–0.3B\textsubscript{1-x}Nd\textsubscript{x}T perovskite crystal structure, the cations Ba\textsuperscript{2+} and Ti\textsuperscript{4+} of BTO can be stably combined to form compounds. But BFO has intrinsic defects due to its chemical element. Firstly, Fe\textsuperscript{3+}/Fe\textsuperscript{2+} leads to a decrease in positive charge. Furthermore, Bi\textsuperscript{3+} is easy to volatilize above 800 °C\cite{32}, which also results the reduction of positive charge. To maintain the electrical neutrality of the crystal structure, more oxygen vacancies are generated. These oxygen vacancies and cationic defects are coupled to form defective dipoles, which increase the conductivity of the ceramic and form leakage currents.\cite{33} High temperatures or strong electric fields accelerate the motion of these defective dipoles and the leakage current becomes more severe. The introduction of Nd\textsuperscript{3+} dopant suppress leakage currents in high temperature.\cite{34} It can provide extra positive charges, a compensatory mechanism to prevent the oxygen vacancies. That disrupts the defect dipole, maintaining high resistivity at high temperature.\cite{35}

\section{Conclusion}

The rare-earth Nd-doped BF-BT ceramics were successfully synthesized via a conventional solid state reaction method. A small amount of Nd\textsuperscript{3+} incorporation significantly suppress leakage current at elevated temperatures. Above 75 °C, Nd\textsuperscript{3+} doped ceramics exhibit lower leakage currents and higher resistivities. This trend intensifies with increasing bias and temperature. The optimum composition (x=0.005) achieved a high d33 172 pC N\textsuperscript{-1} at room temperature, maintaining excellent thermal stability. At 150°C and 1.5kV, the Nd\textsuperscript{3+} doped (x = 0.010) ceramics exhibit a current density reduced by more than 99\% (~10\textsuperscript{-5} A·cm\textsuperscript{-2}) compared to the undoped BF-BT ceramics (~10\textsuperscript{-3} A·cm\textsuperscript{-2}). This indicates that Nd\textsuperscript{3+} doping significantly enhances the leakage resistance of BF-BT ceramics under harsh conditions.









\bibliographystyle{cas-model2-names}

\bibliography{cas-refs}



\end{document}